
\input phyzzx
\singlespace
\twelvepoint
%
%
\font\mathbold=cmmib10 at12pt
\font\bf=cmbx10 at12pt
\newfam\bffam \def\mbf{\fam\bffam\mathbold}
\textfont\bffam=\mathbold
\mathchardef\beta="710C
\mathchardef\gamma="710D
\mathchardef\eta="7111
\mathchardef\xi="7118
\def\mbU{{\mbf U}}
\def\mbV{{\mbf V}}

\def\mbVt{{\tilde \mbV}}

\def\ub{{\bar u}}
\def\vb{{\bar v}}
\def\Ub{{\bar U}}
\def\Vb{{\bar V}}
\def\wb{{\bar w}}
\def\zb{{\bar z}}

\def\bC{{\bf C}}
\def\psib{{\bar \psi}}

\def\Kt{{\tilde K}}
\def\Kh{{\hat K}}
\def\Phit{{\tilde \Phi}}

\def\st{{\tilde s}}
\def\Kahler{K\"ahler\ }
\def\Kahlerian{K\"ahlerian\ }
\def\ra{\partial}
\def\arccoth{{\rm arccoth}}
\def\arctanh{{\rm arctanh}}
\def\const{{\rm const}}

\def\cU{{\cal U}}
\def\PL{{\sl Phys. Lett.}}

\def\IJMP{{\sl Int. J. Mod. Phys.}}
\def\NP{{\sl Nucl. Phys.}}

\def\PR{{\sl Phys. Rev.}}
\def\CMP{{\sl Comm. Math. Phys.}}
\REF\ZU{B. Zumino, \PL  {\bf B87} (1979) 203.}
\REF\SSTVP{Ph. Spindel, A. Sevrin,
W. Troost and A. Van Proeyen,
\NP  {\bf B308} (1988) 662. }
\REF\LR{ U. Lindstr\"om and M.
Ro\v cek, \NP {\bf B222}
(1983) 285.}
\REF\GHR{ S. Gates, C. Hull
and M. Ro\v cek, \NP {\bf
B248} (1984)
157.}
\REF\KKL{ E. Kiritsis,
C. Kounnas and D. L\"ust,
\IJMP{\bf A9} (1994) 1361.}
\REF\CL{ G.L. Cardoso
and D. L\"ust , \PL {\bf B345} (1995) 220.}
\REF\BU{ T. Buscher, \PL {\bf B201} (1988) 466.}
\REF\betafunc{ E. Fradkin
and A. Tseytlin, \NP {\bf B261}
(1985) 1;
C. Callan, D. Friedan,
E. Martinec and M. Perry, \NP {\bf
B262} (1985)
593. }
\REF\BV{G. Bonneau
and G. Valent, {\sl Class. Quant. Grav. }
{\bf 11} (1994) 1133.}
\REF\HW{C. Hull
and E. Witten, \PL  {\bf B160} (1985) 398. }
\REF\RSS{M. Ro\v cek,
K. Schoutens and A. Sevrin,
\PL  {\bf B265} (1991) 303. }
\REF\BLR{T. Buscher,
U. Lindstr\"om and M. Ro\v cek, \PL
{\bf B202} (1988) 94. }
\REF\HEAVENS{ C. Boyer
and J. Finley, {\sl J. Math. Phys.}
{\bf 23} (1982) 1126;
J. Gegenberg and
A. Das, {\sl Gen. Rel. Grav.}
{\bf 16} (1984) 817. }
\REF\TRHEAVENS{ S. Hawking, \PL
{\bf A60} (1977) 81;
G. Gibbons and S. Hawking, \CMP
{\bf 66} (1979) 291;
K. Tod and R. Ward,
{\sl Proc. R. Soc. London Ser.}
{\bf A368} (1979) 411.}
\REF\AXIONIC{ A. Dabholkar,
G. Gibbons, J. A. Harvey
and F. Ruiz-Ruiz,
\NP {\bf B340} (1990) 33;
A. Strominger, \NP {\bf B343} (1990) 167;
C. Callan, J. A. Harvey and A. Strominger,
\NP {\bf B359}  (1991) 611,
{\bf B367} (1991) 60.}
\REF\AXIONICS{ S.J. Rey,
\PR {\bf D43} (1991) 526;
M. Duff and J. X. Lu, \NP
{\bf B354} (1991) 141;
R. Khuri, \NP {\bf B387}  (1992) 315.}
\REF\BS{ I. Bakas, \PL {\bf B343} (1995) 103;
I. Bakas and K. Sfetsos, \PL {\bf B349} (1995) 448}
\REF\HA{S. F. Hassan, CERN-TH-95-98, hep-th/9504148.}
\pubnum={OU-HET 213\cr
hep-th/9505177}
\date{May.1995}
\titlepage
\title{{\bf{\bf Four-Dimensional N=2(4)
Superstring Backgrounds
\vskip .4cm  and
\vskip .4cm  The Real Heavens}}}
\author{Makoto SAKAGUCHI
\foot{e-mail:
gu@funpth.phys.sci.osaka-u.ac.jp}}
\address{\null\hskip-8mm
Department of Physics,
Osaka University, \break
Toyonaka, Osaka 560, JAPAN}
\abstract{We study N=2(4) superstring
backgrounds which are
four-dimensional non-\Kahlerian
with non-trivial dilaton and torsion fields.
In particular we consider
the case that the backgrounds
possess at least one $U(1)$ isometry
and are characterized
by the continual Toda equation
and the Laplace equation.
We obtain a string background
associated with a non-trivial
solution of the continual
Toda equation, which is mapped, under the
T-duality transformation,
to the hyper-\Kahler Taub-NUT instanton background.
It is shown that the integrable property
of the non-\Kahlerian
spaces have the direct origin in the real heavens:
real, self-dual, euclidean, Einstein spaces.
The Laplace equation and
the continual Toda equation imposed on
quasi-\Kahler geometry for
consistent string propagation are
related to the self-duality conditions
of the real heavens
with ``translational'' and ``rotational'' Killing symmetry
respectively.}
\endpage
\chapter{Introduction}

Supersymmetric $\sigma$-models have attracted
attention for various reasons
for a long time.
One of them is their deep relationship
with complex manifold theory.
Recently the interest have been refreshed
in connection with superstring theory.
It has been shown that the number of
supersymmetries
realized on two
dimensional world sheet
restricts the background geometry.
If two dimensional world sheet
theory has $N=1$ supersymmetry,
no-restriction on the background
is imposed.
$N=2$ supersymmetry, however,
imposes a number of conditions.
The simplest case is that of
a torsionless Riemannian background,
which must be a \Kahler manifold
in order to admit $N=2$ supersymmetry[\ZU].
Such $\sigma$-models are
conventionally formulated
in terms of $N=2$ chiral superfields
and the superspace Lagrangian is
just the \Kahler potential.

In the presence of torsion, the situation
becomes considerably complicated.
In this case the background has to admit
two covariantly constant complex
structures.
A typical example of such WZNW
$\sigma$-models are those
with group manifolds as target spaces.
In ref.[\SSTVP] the conditions
for $N=2$ supersymmetry on group manifolds
were found and a complete classification was given.

In ref.[\GHR] it was shown that (2,2)
supersymmetric $\sigma$-models
formulated in terms of chiral and
twisted chiral superfields
describe torsionful target spaces.
Moreover, the abelian T-duality transformation
was formulated
by means of a Legendre transformation
which interchanges a chiral superfield
with
a twisted chiral one in the manifestly
$N=2$ supersymmetry preserving manner.
Then the backgrounds which are dual to
those described by the familiar
(2,2)$\sigma$-models
formulated in terms of chiral superfields
are completely described by
ones formulated in terms
of chiral and twisted chiral superfields.

In order that these geometries provide
consistent string backgrounds
they have to satisfy, adding the dilation field,
the string equations of motion, namely,
the vanishing of $\beta$-functions.
In ref.[\KKL] a systematic discussion
on four-dimensional backgrounds
with $N=2$ world sheet supersymmetry was given.
There a set of conditions were derived,
which are imposed on \Kahler or torsionful
non-\Kahler
with $N=2$ world sheet supersymmetry.
These conditions for consistent
string propagation could be re-expressed
by simple differential equations.
For example a class of non-\Kahlerian backgrounds
including the axionic instanton background
was constructed as solutions
to a simple integrable model i.e.
one with the Laplace equation
as field equation.
Following this line, in the presence of (at least)
one $U(1)$ isometry,
the new four-dimensional non-\Kahlerian background
which has the non-trivial dilaton and torsion fields
was constructed in ref.[\CL].
In this case the constraint
imposed on target space geometry
is related to an integrable model namely one with
the continual Toda equation as field equation and
the relation of the solution
with the hyper-\Kahler Eguchi-Hanson instanton
background was discussed.

In this paper, we explore these lines.
The new superstring background
with non-trivial dilaton and torsion fields,
which is dual to the hyper-\Kahler Taub-NUT
instanton background, is
presented.
The origin of the integrable property
of non-\Kahlerian backgrounds,
which emerge as the Laplace equation
and the continual Toda equation,
is clarified.
It is found that these integrable
equations are related to
those of the real heavens,
which is the self-dual condition
of the Riemann curvature
of the euclidean Einstein gravity.

This paper is organized as follows.

We begin with a review of some
of the relevant aspects
presented in ref.[\LR,\GHR] for
constructing non-trivial four-dimensional
non-\Kahlerian backgrounds
with torsion fields described
by the (2,2) $\sigma$-models
formulated in terms of one chiral
and one twisted chiral superfield.
As is worked out in ref.[\KKL],
adding dilaton field we present
the differential equations
imposed on target space geometry
for consistent string propagation.
Following ref.[\CL] , it is shown that,
due to (at least) one $U(1)$ Killing symmetry,
the condition
imposed on non-\Kahlerian backgrounds
implies the continual Toda equation.
A non-trivial background,
which is dual to the Eguchi-Hanson instanton
background, is obtained as a solution
of the continual Toda equation.
In section 3,
it is found that the non-\Kahlerian background which
is dual to the Taub-NUT instanton background
can be constructed through
a solution of the continual Toda equation.
Section 4 is spent to show that the origin
of integrability lies
in the real heavens.
The last section
is devoted to a summary and discussions.
In the appendix A, the vanishing conditions
of $\beta$-functions
are re-expressed in terms of
the quasi-\Kahler potential and dilaton field.
The duality transformation
by means of a Legendre transformation
is explained in appendix B.

\chapter{The Quasi-\Kahler Geometry
and Integrable Equations}
\section{N=2 Superstring backgrounds}

The most general $N=2$ superspace action
for one chiral superfield $U$\ and
one twisted chiral superfield $V$\
in two dimensions is determined
by a single real function
$K(U,\Ub,V,\Vb)$ [\LR,\GHR]:
$$
S={1\over 2\pi \alpha '}
\int{\rm d}^2xD_+D_-\bar D_+\bar D_-
K(U,\Ub,V,\Vb).\eqn\action\
$$
The superfields $U$ and $V$ obey
a chiral or twisted chiral
constraint
$$
\bar D_\pm U=0,\qquad\bar D_+V=D_-V=0.
\eqn\Constraint
$$
The action \action\ is invariant,
up to total derivatives,
under the quasi-K\"ahler gauge transformations:
$$
K\rightarrow K+\Lambda_1(U,V)+\Lambda_2(U,\Vb)
+\bar \Lambda_1(\Ub,\Vb)+\bar \Lambda_2(\Ub,V).
\eqn\Inva
$$
To read off the target space geometry of the theory
it is convenient
to write down, denoting $u$\ and $v$\
as the lowest component
of the superfield $U$ and $V$ respectively,
the purely bosonic part of the superspace action
\action\ :
$$
S=-{1\over 2\pi\alpha '
}\int{\rm d}^2x\lbrack  K_{u\ub }\partial^a u
\partial_a\ub -K_{v\vb}\partial^a v\partial_a\vb
+\epsilon_{ab}(K_{u\vb}\partial_a u\partial_b\vb
+K_{v\ub}\partial_a v\partial_b\ub)\rbrack,
\eqn\bosonic
$$
where
$$
K_{u\ub}=
{\partial^2K\over\partial U\partial\bar U},\qquad
K_{v\vb}=
{\partial^2K\over\partial V\partial\Vb},\qquad
K_{u\vb}=
{\partial^2K\over\partial U\partial\Vb},\qquad
K_{v\ub }=
{\partial^2K\over\partial V\partial\Ub}.
$$
The target space metric and anti-symmetric tensor
are expressed
in terms of $K$ respectively
$$
G_{\mu\nu}=
\left(
\matrix{0&K_{u\ub}&0&0\cr
K_{u\ub}&0&0&0\cr
0&0&0&-K_{v\vb}\cr
0&0&-K_{v\vb}&0\cr}\right),\qquad
B_{\mu\nu}=
\left(
\matrix{0&0&0&K_{u\vb}\cr
0&0&K_{v\ub}&0\cr
0&-K_{v\ub}&0&0\cr
-K_{u\vb}&0&0&0\cr
}\right).
$$
It follows that the field strength $H_{\mu\nu\lambda}
=\nabla_{\mu} B_{\nu\lambda}+\nabla_{\nu} B_{\lambda\mu}
+\nabla_{\lambda} B_{\mu\nu}$\
can also be expressed entirely in terms
of the function $K$:
$$
H_{u\ub v}
=
-{\partial^3K\over\partial U\partial\Ub\partial V},
\ H_{u\ub\vb}
=+
{\partial^3K\over\partial U\partial\Ub\partial \Vb},
\ H_{v\vb u}
=
+{\partial^3K\over\partial V\partial\Vb\partial U},
\ H_{v\vb\ub }
=-{\partial^3K\over\partial V\partial\Vb\partial
\bar U}.
$$

If $K_{u\ub}$ and $K_{v\vb}$ are positive definite,
the target space possesses $(2,2)$ signature.
To obtain a space with euclidean signature,
we have to require that
they are positive definite
and negative definite respectively.
Note that the metric is non-K\"ahlerian
with torsion,
whereas $N=2$ world sheet supersymmetry
is guaranteed.

Hitherto we have just discussed
the geometrical structure
of the $N=2$ supersymmetric $\sigma$-models.
In string theory,
there is another background field,
namely the dilaton field
$\Phi(u,\ub,v,\vb)$,
so that one adds to the $\sigma$-model action
eqn. \bosonic\  a term of the form ${1\over
2}R^{(2)}\Phi(u,v)$,
where $R^{(2)}$ is the scalar curvature
of the two-dimensional
world sheet.
In order that these backgrounds provide
consistent string solutions, they have to satisfy
the vanishing of the $\beta$-function equations.
Then the requirement of one-loop conformal
invariance of the two-dimensional $\sigma$-model
leads to the following equations of motion
for the background fields [\betafunc],
$$
\eqalign{
&0=\beta_{\mu\nu}^G=R_{\mu\nu}-{1\over
4}H_\mu^{\lambda\sigma}
H_{\nu\lambda\sigma}+
2\nabla_\mu\nabla_\nu\Phi
+O(\alpha'),\cr
&0=\beta_{\mu\nu}^B=
\nabla_\lambda
H_{\mu\nu}^\lambda-
2(\nabla_\lambda
\Phi )H_{\mu\nu}^\lambda+
O(\alpha').}
\eqn\betaf
$$
Moreover,
the vanishing of the dilaton $\beta$-function
is provided
by the equation of motion for dilaton field as
$$
0=\delta c\equiv c-{3D\over 2}=
{3\over 2}\alpha'
\lbrack 4(\nabla\Phi)^2-
4\nabla^2\Phi-R+{1\over 12}
H^2\rbrack+O(\alpha'^{2}).
\eqn\dilaton
$$
In the presence of N=4 world sheet superconformal
symmetry,
the solution to the lowest order
in $\alpha$'
is exact to all orders and $\delta c$
remains zero to
all orders.

The conditions of the vanishing
of $\beta$-functions
were re-expressed in terms of $K$
and $\Phi$ entirely in [\KKL].
There three exclusive cases were considered,
corresponding to the differential equation
which is satisfied by $K$.
In this paper we concentrate ourselves to
two of them, the case(i) and case(ii) explained in
appendix A.
For the case(i), the potential $K$
must satisfy the Laplace equation
$$
K_{u\ub}+K_{v\vb}=0,
\eqn\laplace
$$
and the dilaton field is expressed
in terms of a solution $K$ as
$$
2\Phi=\ln K_{u\ub}+{\rm const.}
\eqn\oisodila
$$
In turn the case(ii) is characterized by
the following nonlinear differential
equation
$$
K_{u\bar u}+K_{ww}e^{K_w}=0,
\eqn\solveb
$$
which determines target space geometry.
The dilaton field is given in terms of a solution
of eqn.\solveb\ to be
$$
2\Phi=\ln K_{ww}+{\rm const.}
\eqn\tisodila
$$
In this case quasi-\Kahler potential
and dilaton field have
$U(1)$ Killing symmetry with respect to
$W$, namely $K=K(u,\ub,w+\wb)
, \Phi=\Phi(u,\ub,w+\wb)$.
We denote the $U(1)$ isometry as $U(1)_w$
for simplicity.

\section{The Hyper-\Kahler Eguchi-Hanson
Instanton and Integrable Equations}

In ref.[\CL] it was shown
that non-\Kahlerian backgrounds
characterized by eqn.\solveb\
arise as a solution of the continual
Toda equation.
Performing the duality transformation
its relation to the
Eguchi-Hanson instanton background
was discussed.

In fact eqn.\solveb\ is re-expressed,
denoting $K_{w}$ as $\cU$ , as
the continual Toda equation
$$
\ra_u\ra_{\ub}\cU+\ra_w^2e^{\cU}=0.
\eqn\contToda
$$
Assuming that $\cU=\ln\alpha(u,\ub)+\ln\beta(w+\wb)$,
eqn.\contToda\ reduces to the Liouville equation
$$
\ra_u\ra_{\ub}\ln\alpha(u,\ub)+k\alpha(u,\ub)=0
\eqn\Liouville
$$
where $\ra_w^2\beta$ is a constant
due to the separation of variables
and is denoted as $ k$.
By using the solution of the Liouville equation
authors of ref.[\CL] employed
the simplest non-trivial solution of
eqn.\contToda\ as
$$
\cU =\ln{-\rho^2+(w+\wb)^2\over {(1+u\ub)^2}}
\eqn\CLgamma
$$
and wrote down
the solution of eqn.\solveb\ as
$$
\eqalign{
K=-2(w+&\wb)+2\rho \arctanh{(w+\wb)\over \rho}\cr
&+(w+\wb)\ln(-\rho^2+(w+\wb)^2)-2(w+\wb)\ln(1+u\ub).
}
\eqn\EHcl
$$
It was shown that after performing the change
$w \rightarrow -w$ the geometry characterized
by \EHcl\ is dualized
with respect to $U(1)_w$ isometry
to give the hyper-\Kahler
Eguchi-Hanson instanton background.
However $w$-sign flipped $K$ is no longer a
solution of eqn.\solveb\ .
In order to make evident the relation
of integrable models
to the hyper-\Kahler Eguchi-Hanson
instanton background,
we present here another solution
of eqn.\solveb\ ,
which is also associated with
the solution \CLgamma\ ,
as
$$
\eqalign{
K=
-2(w+&\wb)-2\rho \arccoth
{-(w+\wb)\over \rho}\cr
&+(w+\wb)\ln(-\rho^2+(w+\wb)^2)-
2(w+\wb)\ln(1+u\ub),}
\eqn\dualEH
$$
which describes the non-trivial
background with torsion;
$$
\eqalign{
&ds^2=
{-4(w+\wb) \over {-\rho^2+(w+\wb)^2}}dwd\wb
-{4(w+\wb)\over {(1+u\ub)^{2}}}dud\ub, \cr
&2\Phi=
\ln|{w+\wb \over{-\rho^2+(w+\wb)^2}}|+
{\rm \const.}\cr
&H_{u\ub w}=
-H_{u\ub \wb}=
{+2 \over {(1+u\ub)^2}}.}
\eqn\dualEHgeo
$$
The scalar curvature
is computed to be given by
$$
R=
{{3\rho^4-14\rho^2
(w+\wb)^2+3(w+\wb)^4}
\over{2(\rho^2-
(w+\wb)^2)(w+\wb)^3}}
$$
which depends only on $w+\wb$
and is asymptotically zero
as $w+\wb\rightarrow \pm\infty$.

Let us comment on the difference
of eqn.\dualEH\ from eqn.\EHcl\ .
There appears \
$-\arctanh$\ ${(w+\wb)\over \rho}$ \
instead of \
$\arccoth {-(w+\wb)\over \rho}$\ .
Since the geometrical objects are
expressed in terms of
the derivatives of $K$,
both potentials describe the same geometry
except for the range of $(w+\wb)^2$,
which correspond to describing
two different coordinate patches.
For the solution \EHcl\ and \dualEH\
the range must be $(w+\wb)^2 <\rho^2$
and $(w+\wb)^2 >\rho^2$ respectively.

There exist different backgrounds
with those discussed so far,
which are called dual background
and obtained by duality transformation.
Now we dualize the solution \dualEH\
with respect to $U(1)$ isometry with respect to
the twisted chiral superfield
$W$ following the procedure
described in the appendix B.
The duality transformation interchanges
a twisted chiral superfield $W$ with
a chiral superfield $\Psi$ so that the dual theory
is described by two chiral superfield.
Then dual geometry is \Kahler.
The dual \Kahler potential is determined by
$$
\Kt=K-(w+\wb)(\psi+\psib)
$$
with a constraint equation $0=K_w-(\psi+\psib)$
which determines $(w+\wb)$ in terms of
$(\psi+\psib)$ and $u,\ub$.
\foot{We use small letters $w,\ z,\ \psi,\ ...$
for the lowest component
of superfields $W,\ Z,\ \Psi,\ ...$}
Now the independent variables are $\psi$ and $u$.

The constraint is computed to be
$$
(w+\wb)^2=\rho^2+e^{\psi+\psib}(1+u\ub)^2
\eqn\constraint
$$
which implies $(w+\wb)^2 > \rho^2$.
This is compatible only when
the potential is of the form \dualEH .
The constraint \constraint\ implies that
$w+\wb = \pm \sqrt{\rho^2+
e^{\psi+\psib}(1+u\ub)^2}$.
We use here $w+\wb<0$
as a solution of \constraint\
in order that the geometry \dualEHgeo\
has euclidean signature.

The dual \Kahler potential is computed to be
$$
\Kt=2g-2\rho \arccoth{g \over \rho},\quad\quad
g=\sqrt{\rho^2+e^{\psi+\psib}(1+u\ub)^2}.
\eqn\EH\
$$
Under the coordinate transformation
$$
z^1=e^{\psi/2},\quad z^2=e^{\psi/2}u
$$
the dual \Kahler potential \EH\ describes
hyper-\Kahler Eguchi-Hanson metric [\BV] ;
$$
g_{1\bar 1}=
4({g \over Q^2}|z^2|^2+
{1 \over g}|z^1|^2),\qquad
g_{1\bar 2}=
-4{\rho^2 \over {gQ^2}}z^2\bar z^1,
\qquad
g_{2\bar 2}=
4({g \over Q^2}|z^1|^2+
{1 \over g}|z^2|^2),
\eqn\EHmetric
$$
where $Q=|z^1|^2+|z^2|^2$
and $g=\sqrt{\rho^2+Q^2}$.
In other words the hyper-\Kahler
Eguchi-Hanson instanton background
is dual with respect to $U(1)_\psi$ isometry,
namely overall $U(1)$ isometry,
to the quasi-\Kahler background
\dualEHgeo\ .

\chapter{The Hyper-\Kahler Taub-NUT
Instanton and Integrable Equations}

We consider in this section the relation of
the hyper-\Kahler Taub-NUT instanton
to the integrable model of the
quasi-\Kahler geometry.
It is shown that the hyper-\Kahler
Taub-NUT instanton background
is also dual
to a quasi-\Kahler geometry characterized by
a solution of the continual Toda equation.

Let us first consider a solution
of the continual Toda equation \contToda\
with $(u,\ub)$ denoted as $(z,\zb)$
to avoid confusion in the following discussion;
$$
\cU =2\ln[-4(w+\wb)\cosh^{-1} 4(z+\zb)].
\eqn\dualTNflatgamma
$$
The quasi-\Kahler potential is
$$
K=-2(w+\wb)+2(w+\wb)\ln
[-4(w+\wb)\cosh^{-1} 4(z+\zb)].
\eqn\dualflatTN
$$
The corresponding geometry is the same as
\dualEH\ with $\rho=0$.
In fact the quasi-\Kahler potential
\dualEH\ with $\rho=0$ can be
transformed to \dualflatTN\ under
the quasi-\Kahler transformation
as follows.
The potential \dualEH\ with $\rho=0$
is expressed as
$$
K=-
2(w+\wb)+2(w+\wb)\ln[-(w+\wb)]-2(w+\wb)
\ln[1+u\ub].
$$
We perform the coordinate transformation
$u=e^{8z}$ but
the resulting potential
is not a solution of \solveb\ .
In order to obtain a solution
the quasi-\Kahler transformation
(A.15) of the form
$K\rightarrow K+8(w+\wb)(z+\zb)$
and (A.16) of the form
$K\rightarrow K+(w+\wb)2\ln8$ must be followed.
Under these transformations
we obtain eqn.\dualflatTN\ as
a solution of \solveb\ .

Next let us dualize \dualflatTN\
with respect to $U(1)_w$ isometry.
We obtain the dual \Kahler potential,
interchanging a twisted chiral superfield $W $
with a chiral superfield $Z^1$ , as
$$
\Kt=
{1\over 2}e^{{1\over 2}
(z^1+\bar z^1)}\cosh{1\over 2}
(z^2+\bar z^2)
\eqn\flatTN
$$
where we introduce $z^2\equiv 8z$.
The corresponding geometry
is completely flat but
the coordinate system of eqn.\flatTN\
suggests the relation
to the hyper-\Kahler Taub-NUT metric
as is seen below.

In order to see the relation of
the \Kahler potential \flatTN\ to
the hyper-\Kahler Taub-NUT metric,
let us employ the following expression
for the hyper-\Kahler
Taub-NUT potential [\BV]
$$
K^{TN}
={s \over 2}[1+{1\over 4}
\lambda s(1+\cos^2\theta)]
\eqn\TN
$$
with the \Kahler coordinate
$$
\eqalign{
z^1&=-i\phi+\ln(s\sin\theta)\cr
z^2&=i\psi-\ln(\tan{\theta \over 2})
+\lambda s\cos\theta}
\eqn\kahlerco
$$
where $\lambda$ corresponds to the magnetic mass.
The variables $\theta, \phi\ {\rm and}\ \psi$ are
angular ones in the polar coordinates
with the range $0\le \theta \le \pi,
0\le \phi \le 2\pi\
{\rm and}\ 0\le \psi \le 4\pi$.
The $s$ is
the square of the radial variable.
The corresponding \Kahler metric
in the coordinate \kahlerco\
is given by
$$
g_{1\bar 1}=
{s\over 4}\big({\cos^2\theta \over{1+\lambda s}}
+(1+\lambda s)\sin^2\theta \big),\quad
g_{1\bar 2}=
{s\over 4}{\cos\theta \over{1+\lambda s}},\quad
g_{2\bar 2}=
{s\over 4}{1 \over{1+\lambda s}}.
\eqn\TNmetric
$$
These clarify the relation
of the eqn.\flatTN\ to the eqn.\TN\ as
$$
\Kt={s \over 2}=K^{TN}\mid_{\lambda=0}\ .
$$
In this way we see that
the quasi-\Kahler potential \dualflatTN\
which is solved via the continual Toda equation
is dual to
the hyper-\Kahler Taub-NUT metric
with zero magnetic mass.
{}From this fact a question arises whether
there also exist the relation of
the continual Toda equation to
the hyper-\Kahler Taub-NUT metric
with $\lambda \not=0$.
We show below that this is the case.

Since we have a hyper-\Kahler
Taub-NUT metric associated with the
\Kahler potential \TN\
in the coordinate \kahlerco\
which is dual to a solution of \solveb\
in the case of $\lambda =0$,
we try to dualize \TN\
with respect to $U(1)_{z^1}$ isometry.
Due to the ${\bf Z}_2$ property
of duality transformation,
we can construct the \Kahler potential \TN\
from the resulting quasi-\Kahler potential.

Here we deal with the case that the
\Kahler geometry which is
described by two chiral superfields
is dualized to
the quasi-\Kahler geometry described
by a chiral
and a twisted chiral superfield.
This causes the modification
of the duality transformation
(B.2) and (B.3) as follows.
The dual quasi-\Kahler potential,
interchanging a chiral superfield $Z^1$
with a twisted chiral superfield $W$,
is determined by
$$
K=K^{TN}+(z^1+\bar z^1)(w+\wb),
$$
with a constraint $0=\ra_{z^1}K^{TN}+(w+\wb)$.
The constraint becomes
$(w+\wb)=
-{1\over 4}s(1+{1\over 2}\lambda s \sin^2\theta)$
and we obtain the quasi-\Kahler potential
$$
K=
{s\over 2}[1+{1\over 4}\lambda
s (1+\cos^2\theta)
-(1+{1\over 2}\lambda s
\sin^2\theta)\ln(s\sin\theta)],
\eqn\dualTN
$$
where the quasi-\Kahler coordinate is given,
denoting $z^2$ as $8z$, by
$$
\eqalign{
z&=
{1\over 8}\big(i\psi-\ln(\tan{\theta \over 2})+
\lambda s \cos\theta\big),\cr
w&=
{i\over2}\phi-{1\over 8}s(1+{1\over 2}
\lambda s \sin^2\theta).}
\eqn\dualkahlerco
$$
In order to show that the quasi-\Kahler
potential \dualTN\
arises as the solution
of the continual Toda equation,
we compute the derivatives
of $K$ to be given by
$$
\eqalign{
&K_z=
2 s\cos\theta,\cr
&K_w=
2\ln(s\sin\theta),\cr
&K_{z\zb}=
8s\sin^2\theta
\big({\cos^2\theta \over {1+\lambda s}}+
(1+\lambda s )\sin^2\theta\big)^{-1},\cr
&K_{w\wb}=
-{8\over s}
\big({\cos^2\theta \over {1+\lambda s}}+
(1+\lambda s )\sin^2\theta\big)^{-1},\cr
&K_{z\wb}=
-8{\cos\theta\over{1+\lambda s}}
\big({\cos^2\theta \over {1+\lambda s}}+
(1+\lambda s )\sin^2\theta\big)^{-1}.
}
\eqn\deri
$$
One can easily see that eqns.\solveb\
and \contToda\ are satisfied.
Thus the potential \dualTN\ describes
$N=2$ superstring background.
The corresponding geometry which
is associated with the potential
\dualTN\ is given by
$$
\eqalign{
ds^2&=
{16\over s}
\big({\cos^2\theta \over {1+\lambda s}}+
(1+\lambda s )\sin^2\theta\big)^{-1}
\big(dwd\wb+s^2\sin^2\theta dzd\zb\big),\cr
H_{z\zb w}&=
-H_{z\zb \wb}=
+32\sin^2\theta
\big({\cos^2\theta \over {1+\lambda s}}+
(1+\lambda s )
\sin^2\theta\big)^{-3}\cr
&\quad\quad\quad\quad\quad
\times\big(\sin^2\theta+
{\cos^2\theta \over {1+\lambda s}}
+{2\lambda s \over{(1+\lambda s)^2}}-
{\lambda^2 s^2\sin^2\theta
\over{(1+\lambda s)^3}} \big),\cr
H_{w\wb z}&=-H_{w\wb \zb}=
-32\lambda\sin^2\theta\cos\theta
\big({\cos^2\theta \over {1+\lambda s}}+
(1+\lambda s )\sin^2\theta\big)^{-3}\cr
&\quad\quad\quad\quad\quad\times
\big({3\over {1+\lambda s}}-
{\lambda^2 s^2\cos^2\theta
\over{(1+\lambda s)^3}}\big),\cr
2\Phi&=-\ln{s}-
\ln\mid {\cos^2\theta \over {1+\lambda s}}+
(1+\lambda s )\sin^2\theta\mid+{\rm const}.}
\eqn\geo
$$
In the real coordinate the metric is expressed as
$$
\eqalign{
ds^2=
&{1+\lambda s\over
{4s}}(ds^2+s^2d\theta^2)\cr
&+{1\over {4s}}
\big({\cos^2\theta \over {1+\lambda s}}+
(1+\lambda s )\sin^2
\theta\big)^{-1}
(16d\phi^2+s^2
\sin^2\theta d\psi^2),
}
$$
where $\phi$ and $\psi$ are introduced in
\dualkahlerco\ to be the imaginary part
of $w$ and $z$ respectively.
One find that in the case $\lambda=0$
the metric is dual with
respect to rotational $U(1)$
isometry associated
with the angular coordinate $\phi$
to flat metric as expected.
The corresponding scalar curvature is
$$
\eqalign{
R=&(24+176\lambda s+
312\lambda^2 s^2+52\lambda^3 s^3
-235\lambda^4 s^4-
168\lambda^5 s^5-
32\lambda^6 s^6\cr
&+144\lambda s\cos2 \theta+
488\lambda^2 s^2 \cos2 \theta
+592\lambda^3 s^3 \cos2 \theta+
388\lambda^4 s^4 \cos2 \theta \cr
&+160\lambda^5 s^5 \cos2 \theta+
32\lambda^6 s^6 \cos2 \theta
-4\lambda^3 s^3 \cos4 \theta+
7\lambda^4 s^4 \cos4 \theta \cr
&+8\lambda^5 s^5 \cos4 \theta)
\big{/}\big(4s(1 + \lambda s)^5
({\cos^2\theta \over 1+\lambda s} +
\sin^2\theta (1+\lambda s))^2\big).
}
$$It follows that the scalar curvature
is asymptotically zero
($s\rightarrow +\infty$).
For the case $\lambda =0 $ ,
the curvature singularity is at $s=0$.
In turn if we consider $\lambda < 0$,
there are two singularities at $s=0$
and $-1/\lambda$.

We obtained the non-trivial
superstring background
which is a solution of eqn.\solveb\
and is dual to the hyper-\Kahler Taub-NUT solution.
As is mentioned in the section 2 the eqn.\solveb\ is
related to the continual Toda equation,
so that a class of the solution
of the continual Toda equation
emerge.
The corresponding solution
of the continual Toda equation \contToda\ :
$\ra_z\ra_{\zb}\cU+\ra_w^2e^\cU=0$
is
$$
\cU =2\ln(s\sin\theta),
\eqn\TNgamma
$$
where the r.h.s. is expressed implicitly in terms of
$z$ and $w$ by means of eqn.\dualkahlerco\ .
If the term $\ra_z\ra_\zb \cU $
is independent of the variable $w$:
$\ra_w(\ra_z\ra_\zb \cU )=0$,
the continual Toda equation can be reduced
to the Liouville equation.
We find that the solution \TNgamma\
is not the case and
can not be reduced to the one
of the Liouville equation.

\chapter{The Quasi-\Kahler Geometry
and The Real Heavens}
In section 2 and 3 we present
$N=2$ superstring quasi-\Kahler
backgrounds which are dual to Eguchi-Hanson
instanton and
Taub-NUT instanton respectively.
Both are obtained through solutions
of the continual Toda equation.
In this section we study the origin
of the integrable property
of the quasi-\Kahler geometry
for the case(i) and case(ii).
It is shown that the integrability
can be understand
as a direct reflection of
the one of the real heavens:
real, self-dual, euclidean,
Einstein spaces.
It was shown, in ref.[\HEAVENS],
that all solutions to the real vacuum
Einstein equations
with self-dual or anti-self-dual curvature:
$$
R_{\mu\nu\rho\sigma}=
\pm {1\over 2}\sqrt{\det G}\
{\epsilon_{\rho\sigma}}^{\kappa\lambda}
R_{\mu\nu\kappa\lambda},
$$
in the presence of at least
one Killing symmetry,
fall into two cases which correspond
to two distinct
types of Killing vectors.
The first type, what is called
``translational'', corresponds to
Killing vectors $K_{\nu}$ with self-dual
or anti-self-dual
covariant derivatives
$$
\nabla_{\mu}K_{\nu}=
\pm {1\over 2}\sqrt{\det G}\
{\epsilon_{\mu\nu}}^{\kappa\lambda}
\nabla_{\kappa}K_{\lambda}.
$$
The second type, what is called
``rotational'', includes all other
Killing vectors.

These gravitational backgrounds
are hyper-\Kahler
and consistent with $N=4$ world sheet
supersymmetry.
The relation of the
world sheet supersymmetry to
T-duality transformation is
recently considered in
ref.[\BS] by using these backgrounds.

In the following we show that
the quasi-\Kahler backgrounds for
the case(i) and case(ii) are dual
to the real heavens
with (at least) one ``translational''
Killing symmetry
and ``rotational'' one respectively.

At first we consider the case(i).
The string backgrounds admit a conformally
flat metric coupled to
axionic instanton and have been considered
before[\KKL,\AXIONIC,\AXIONICS].
To make the statement concrete
we recall the quasi-\Kahler backgrounds:
$$
\eqalign{
&ds^2=
2K_{u\ub}dud\ub-2K_{v\vb} dvd\vb=
2K_{u\ub}(dud\ub+dvd\vb),\cr
&H_{u\ub v}=-\ra_u \ra_\ub \ra_v K,
\qquad
H_{u\ub \vb}=
+\ra_u \ra_\ub \ra_\vb K, \cr
&H_{v\vb u}=
+\ra_v \ra_\vb \ra_u K,
\qquad
H_{v\vb \ub}=
-\ra_v \ra_\vb \ra_\ub K,\cr
&2\Phi=
\ln K_{u\ub}+
{\rm const.}=
\ln(-K_{v\vb})+
{\rm const.}}
\eqn\duallaplace
$$
with $K$ satisfying
the flat Laplace equation
$(\ra_u \ra_\ub+
\ra_v \ra_\vb )K=0$.
If there exist $U(1)_v$
isometry ($U(1)_u$ isometry),
denoting the Killing vector
as $\ra/\ra \tau$,
the real coordinates
$(\tau,x,y,z)$ are introduced by
$$
v=z+i\tau,\quad u=x+iy \qquad
( u=z+i\tau,\quad v=x+iy ).
$$
In the following we distinguish
these cases by means of
upper (lower) sign.
In these coordinates eqns.\duallaplace\
are expressed by the following
$$
\eqalign{
&ds^2=
g_{\tau\tau}(d\tau+A_idx^i)^2 +
{\bar g}_{ij}dx^i dx^j,\cr
&g_{\tau\tau}=
2K_{u\ub},
\quad {\bar g}_{ij}=2K_{u\ub}\delta_{ij},
\quad A_i=0,\cr
&H_{\tau xy}=
\mp2\ra_z K_{u\ub},
\quad
H_{\tau yz}=\mp2\ra_x K_{u\ub},
\quad
H_{\tau zx}=\mp2\ra_y K_{u\ub},\cr
&2\Phi=
\ln K_{u\ub} +{\rm const.}
}
\eqn\realduallaplace
$$
Denoting $2K_{u\ub}$ as $V$,
the anti-symmetric tensor
$B_{\mu\nu}$ can be chosen as
$B_{\tau i}=\omega_i$
with satisfying the special condition:
$\nabla V=\pm\nabla \times \omega$.

Now we perform the duality transformation
with respect to $\tau $ direction:
$$
\eqalign{
&{\tilde g}_{\tau\tau}=1/g_{\tau\tau},
\quad {\tilde A_i}=B_{\tau i},
\quad {\tilde B}_{\tau i}=A_i,\cr
&{\tilde B}_{ij}=B_{ij}-2A_{[i}B_{j]\tau},\cr
&2\Phit=2\Phi-\ln g_{\tau\tau},
\quad {\tilde {\bar g}}_{ij}={\bar g}_{ij}.}
\eqn\realduality
$$
The resulting dual backgrounds are given by
$$
d\st^2={1\over V}(d\tau +\omega_i dx^i)^2
+V(dx^2+dy^2+dz^2)
\eqn\heaventranslational
$$
where $\omega_i$ are constrained
to satisfy the condition
$$
\nabla V=\pm\nabla \times \omega,
\eqn\lap
$$
hence $V$ satisfys the flat Laplace equation.
It was shown in ref.[\HEAVENS] that,
in the presence of (at least) one
``translational'' Killing symmetry,
solutions to the real vacuum Einstein equations
with self-dual or anti-self-dual curvature
are completely determined
by $V$ satisfying the condition \lap\ with
metric \heaventranslational\ [\TRHEAVENS].
Localized solutions of the flat Laplace equation
correspond to
the multi-asymptotically locally
euclidean instantons or
the multi-Taub-NUT instantons,
depending on the asymptotic nature.

We next consider the case(ii).
The quasi-\Kahler backgrounds
have the following form,
denoting
$ K_{w}=
\ra_{w+\wb}K \equiv
{\cU (u,\ub,w+\wb)} $,
$$
\eqalign{
&ds^2=
-2\ra_{w+\wb}\
\cU dwd\wb -2\ra_{w+\wb}\
\cU e^{\cU}dud\ub,\cr
&H_{u\ub w}=
-H_{u\ub \wb}=
-\ra_u \ra_\ub\cU,
\quad H_{w\wb u}=
\ra_{w+\wb}\ra_u\cU,
\quad H_{w\wb \ub}=
-\ra_{w+\wb}\ra_\ub\cU,\cr
&2\Phi=
\ln\ra_{w+\wb}\cU,
}
\eqn\qK
$$
with $\cU$ satisfying the continual Toda equation
$\ra_u \ra_\ub \cU+
\ra_{w+\wb}^2\ e^\cU=0$.
It is convenient to introduce
the real coordinates $(\tau,x,y,z)$
as
$$
w=-(z+i\tau),\quad u=
\cases{y+ix,\cr x+iy. \cr}
\eqn\realco
$$
We denote the upper and lower case
to correspond to the upper and lower sign
in the following.
In these coordinates eqns.\qK\
are expressed by
$$
\eqalign{
&ds^2=
g_{\tau\tau}(d\tau+A_idx^i)^2
+ {\bar g}_{ij}dx^i dx^j,\cr
&g_{\tau\tau}=\ra_z \cU ,
\quad A_i=0,\cr
&{\bar g}_{ij}=
{\rm diag}(\ra_z \cU e^\cU ,
\ra_z \cU e^\cU , \ra_z \cU),\cr
&H_{\tau xy}=
\mp(\ra_x^2+\ra_y^2)\cU,
\quad H_{\tau xz}=
\mp\ra_z\ra_y \cU,
\quad H_{\tau yz}=
\pm\ra_z\ra_x \cU,\cr
&2\Phi=
\ln\ra_z\cU +
{\rm const.}
}
\eqn\realqK
$$
with $\cU$ satisfying $(\ra_x^2+ \ra_y^2)
\cU+\ra_z^2\ e^\cU=0$.
The torsion fields are compatible
with choosing the anti-symmetric
tensor $B_{\mu\nu}$ as
$$
B_{\tau x}=\mp\ra_y \cU,\quad
B_{\tau y}=\pm\ra_x\cU,
$$
and the other components are zero.

Now the Killing vector corresponding
to the $U(1)_w$ isometry is $\ra/\ra \tau$
and we perform the duality transformation
\realduality\ .
The resulting dual backgrounds are given by
$$
d\st^2=
{1\over \ra_z\cU}(d\tau \mp\ra_y
\cU dx\pm\ra_x \cU dy)^2
+\ra_z\cU
[e^\cU (dx^2+dy^2) +dz^2]
\eqn\heaven
$$
with $\cU$ satisfying the continual Toda equation
$$
(\ra_x^2+ \ra_y^2) \cU+\ra_z^2\ e^\cU=0.
\eqn\contToda
$$
It was shown in ref.[\HEAVENS] that,
in the presence of (at least)
one ``rotational'' Killing symmetry,
solutions to the real vacuum Einstein
equations with
self-dual or anti-self-dual curvature
are completely determined
by $\cU$ satisfying eqn.\contToda\
with the metric \heaven\ .
In the above, we considered the case
that the quasi-\Kahler backgrounds
have one $U(1)$ isometry with respect
to a twisted chiral superfield.
If we consider the case $C_1=0,C_2\not= 0$
instead of
the case(ii) $C_1\not=0,C_2= 0$ in section 2,
the quasi-\Kahler backgrounds turn to possess one
$U(1)$ isometry with respect to a chiral superfield.
In this case, denoting $w$ and $u$ as
the lowest component of
a chiral and twisted chiral superfield respectively,
the metric and torsion fields $H_{\mu\nu\rho}$
have opposite sign to eqns.\qK\ .
Introducing the real coordinates \realco\
with the change $w\rightarrow -w$,
the dual backgrounds have
the metric \heaven\ with the constraint \contToda\
again.

As a consequence, we can state that
the origin of the integrability
of the quasi-\Kahlerian for
the case(i) and case(ii)
lies in the real heavens.

In section 2 and 3, the quasi-\Kahler
backgrounds which are dual to
the Eguchi-Hanson and Taub-NUT
instanton background respectively
are constructed.
Since these instanton backgrounds
admit not only
``translational'' Killing symmetry
but also ``rotational'' one,
they can be written in the form \heaven\ .
The multi-ALE and multi-Taub-NUT
instanton backgrounds don't
admit additional ``rotational''
Killing symmetry in general
except for the Eguchi-Hanson and
Taub-NUT instanton backgrounds.
Hence the quasi-\Kahler backgrounds
which are dual to
these instantons can not be constructed
for the case(ii).

\chapter{Summary and Discussions}
In this section, we first summarize
our result and then briefly
discuss their generalizations.

We investigate four dimensional
$N=2$ superstring backgrounds
which are described
by a chiral superfield and a
twisted chiral one.
In particular we considered the case
where there is (at least) one Killing
symmetry and the quasi-\Kahler
potential is determined
by the continual Toda equation.
We found that the background which
is dual to
the well-known Taub-NUT instanton
background arises through
a non-trivial solution to
the continual Toda equation.
We clarify the relationship
of the quasi-\Kahler
backgrounds with the
real heavens i.e.
the real, self-dual, euclidean,
Einstein spaces.
It is found that the quasi-\Kahler
backgrounds
for the case(i) and (ii)
are dual to the real heavens with a
``translational''
Killing symmetry and a ``rotational''
one respectively.
Then it was found that the origin of
the integrable property
lies in the real heavens.

Since the hyper-\Kahler Taub-NUT and
Eguchi-Hanson instanton background
is known to be consistent with $N=4$
world sheet supersymmetry,
we may expect that the corresponding
quasi-\Kahler backgrounds
are exact to all orders of $\alpha'$
if the duality transformation
preserve $N=4$ world sheet supersymmetry.
In ref.[\BS, \HA] the relation of world
sheet supersymmetry and
T-duality transformation is considered.

Four-dimensional $N=2$ superstring backgrounds
described by the (2,2) $\sigma$-models
formulated in terms of a chiral
and a twisted chiral superfields
include another case than the
case(i) and (ii) studied
in this paper.
For the case(iii),
where there are at least two $U(1)$
Killing symmetry
and the quasi-\Kahler potential is determined
by a non-linear differential
equation, we find several interesting
backgrounds.
One of them is a class of conformally
flat backgrounds
with (2,2) signature
which have non-trivial dilaton
and torsion fields.
The dual backgrounds are hyper-\Kahler
nevertheless possesses linear
dilaton field.
Another interesting class
is the direct products
of two 2-dim backgrounds.
The product of
$SU(2)/U(1)
\otimes SL(2,{\bf R})/U(1)$
which is known to possess
$N=4$ world sheet supersymmetry
falls into this class.
The case(iii) covers
these non-trivial solutions
but the general
solution is unknown.

We studied (2,2) $\sigma$-models
described
in terms of twisted chiral and chiral
superfields.
It is known that these $\sigma$-models
put a strong
restriction on the
background geometry, namely,
two complex structures
must ${\it commute}$.
The two ${\it commuting}$
complex structures emerge
when one consider the WZNW
$\sigma$ models only on
$SU(2)\otimes U(1)$ or $U(1)^4$
among the various
group manifolds[\RSS].
Thus the generic (2,2)
supersymmetric $\sigma$-models
can not be exhausted
employing these superfields.

In ref.[\BLR] the (2,2) $\sigma$-models
formulated
in terms of semi-chiral superfields,
which satisfy only
a left-handed or right-handed
chirality condition
but not both simultaneously,
are shown to possess two
non-commuting complex
structures and correspond to
the generic case.
So far only the case of (2,2)
world sheet supersymmetry
has been considered.
If heterotic (2,0) $\sigma$-models
are considered,
the metric and torsion are given
by a complex vector
potential[\HW].
The two dimensional action is
formulated in terms of
(left-handed or right-handed) chiral
superfields in which
the vector potential appears.

It is intriguing problem for us
to investigate the backgrounds
which are described by these
$\sigma$-models
in the string context.
\vskip 1cm
\centerline{\bf Acknowledgments}
We would like to thank Prof.
H. Kunitomo
and Prof. H. Itoyama for helpful
suggestions and encouragement.

\centerline{\bf APPENDIX.A}
In this section we consider
the vanishing conditions of
$\beta$-functions imposed on $K$.
The equations to be solved
are ten for $\beta_{\mu\nu}^G$,
six for $\beta_{\mu\nu}^B$ and one for
$\beta^\Phi$.
These were re-expressed,
defining $\mbU =\log K_{u\ub}$
and $\mbV =\log K_{v\vb}$,
in terms of $\mbU ,\mbV$
and $\Phi$ entirely in ref.[\KKL].

It follows from the six vanishing
conditions of $\beta_{uu}^G$ ,
$\beta_{uv}^G$ ,
$\beta_{u\vb}^G$ , $\beta_{vv}^G$ ,
$\beta_{uv}^B$ and $\beta_{u\vb}^B$
that
$$
\eqalign{
&\partial_u {\mbV} =
2\partial_u\Phi +{\bar \bC_1}(\ub)
\exp{\mbU} ,\cr
&\partial_v {\mbU} =
2\partial_v\Phi +{\bar \bC_2}(\vb)
\exp{\mbV} ,}
\eqno(A.1)
$$
where $\bar \bC_1(\ub)$ and $\bar
\bC_2(\vb)$ are arbitrary
anti-holomorphic functions.
The complex conjugate of
the above six vanishing conditions tell
us that analogous equations with
$\bC_1(u)$ and $\bC_2(v)$
hold for the derivatives
with respect to $\ub$ and $\vb$ .

The remaining five conditions
are the vanishing of
$\beta_{u\ub}^G$ , $\beta_{v\vb}^G$ ,
$\beta_{u\ub}^B$ , $\beta_{v\vb}^B$
and $\beta^\Phi$.
In order to proceed with taking into
account of
these conditions,the following
three exclusive cases were considered:
(i) $\bC_1=\bC_2=0$, (ii) $\bC_1=0$,
$\bC_2\neq 0$, (iii) $\bC_1,\bC_2\neq 0$.

In the following we concentrate ourselves
to the case(i) and case(ii).

For the case(i), the conditions \betaf\
are re-expressed by
the following set of differential
equations:
$$
\eqalign{
&\partial_u(\mbV-2\Phi)=
\partial_{\bar u}
(\mbV-2\Phi)=0,
\quad\partial_v\partial_\vb
(\mbV-2\Phi)=0,\cr
&\partial_v(\mbU-2\Phi)=
\partial_\vb (\mbU-2\Phi)=0,
\quad\partial_u
\partial_{\bar u}
(\mbU-2\Phi)=0,
}
\eqno(A.2)
$$
which can be solved by
$\mbU-2\Phi={\rm const.}$
and $\mbV-2\Phi={\rm const.}$
The potential $K$ must satisfy
$$
K_{u\ub}=K_{v\vb}e^c
\eqno(A.3)
$$
where $c$ is a constant.
The dilaton field is expressed as
$$
2\Phi=\ln|K_{u\ub}|+{\rm const.}
\eqno(A.4)
$$
Without any loss of generality,
we can consider the constant $c$
is pure imaginary
since the real part of it can
be absorbed by rescaling $v$.
Moreover, it is restricted to
$0$ or $i\pi$
for the nontrivial solution.
For the case $c=0$
the corresponding backgrounds have
(2,2) signature.
To obtain euclidean backgrounds
we must choose $c=i\pi$.
In this case, the eqn.(A.3)
is nothing but the Laplace equation.

For the case(ii),
it is very useful to perform
the following
change of coordinates [\KKL]:
$$
w=\int{{\rm d}v\over \bC_2(v)}.
\eqno(A.5)
$$
Combining a remaining condition
of $\beta_{u\ub}^B=0$ with
eqns.(A.1) we obtain that
$\Phi=\Phi(u,\ub,w+\wb)$,
$\mbV=\mbV(u,\ub,w+\wb)$
and $\mbU=\mbU(u,\ub,w+\wb)$.
Thus this case necessarily leads to at
least one $U(1)$ isometry.
The sixteen conditions for \betaf\
are entirely re-expressed
by the following set of
differential equations:
$$
\eqalignno{
&\partial_u(\mbVt-2\Phi)=
\partial_{\bar u}(\mbVt-2\Phi)=0 ,
&(A.6)\cr
&\partial_w(\mbU-2\Phi)=
\exp\mbVt ,
&(A.7)\cr
&\partial_w^2(\mbVt-2\Phi)=0 ,
&(A.8)\cr
&\partial_u\partial_{\bar u}
(\mbU-2\Phi)=\partial_w
\exp\mbU ,&(A.9)}
$$
where we denote
$\mbVt=\ln K_{ww}=
\ln(\bC_2(v)\bar\bC_2(v)
K_{v\vb})$.
The last two equations
follow from the vanishing of
$\beta_{u\bar u}^G$ and
$\beta_{v\bar v}^G$.
The condition
$\beta_{v\vb}^B=0$
leads no additional condition.
Eqns.(A.6) and\ (A.8)
can be solved to give
$$
\mbVt-2\Phi=
c_1(w+\bar w)+
c_2,\eqno(A.10)
$$
where $c_1$ and $c_2$
are integration constants.
Integrating (A.7)
with respect to $w$
and (A.9) with respect to
$u$ and $\ub$ lead to
the equation:
$\ln K_{u\ub}=
2\Phi+K_w+
\lambda(u)+{\bar \lambda}(\ub)$
where $\lambda(u)$ and
${\bar \lambda}(\ub)$ arise as
integration ``constants''.
Eliminating $\Phi$ by using eqn.(A.10),
the potential $K$ has to
satisfy
$$
K_{ww}=
K_{u\bar u}e^{-K_w+c_1(w+\bar w)+
c_2-\lambda(u)-{\bar \lambda}(\ub).
}
\eqno(A.11)
$$
The terms
$\lambda(u)$\ and\ ${\bar \lambda}(\ub)$
can be absorbed by the
holomorphic coordinate transformation
$$
e^{\lambda(u)}du=
dF,\quad
e^{{\bar \lambda(\ub)}}d\ub=
d{\bar F},
\eqno(A.12)
$$
which transform
the eqn.(A.11) to
$$
K_{ww}=K_{u\bar u}e^{-K_w+c_1(w+\bar w)+c_2},
\eqno(A.13)
$$
where we denote
$(F,{\bar F})$ as $(u,\ub)$ again.
Eqn.(A.13) determines target
space geometry.
The dilaton field is given
in terms of a solution
of eqn.(A.13) to be
$$
2\Phi=
\ln K_{w\wb}-
c_1(w+\wb)-c_2.
\eqno(A.14)
$$
Using the above expressions
(A.13) and (A.14)
one can find that the central
charge deficit $\delta c$
is proportional to the
constant $c_{1}$:
$\delta c=-3\alpha' c_1$ and
thus we set $c_1=0$ from now on.
As a consequence,
in order that one-loop $\beta$-functions
\betaf\ and \dilaton\ vanish,
for case(ii) it is the differential
equation (A.13)  with $c_1=0$
that determine the string background.
The dilaton field $\Phi$
is determined by the solution
of (A.13) with $c_1=0$
through eqn.(A.14).

The (quasi-)K\"ahler
transformation is
a gauge transformation
for the spacetime geometry,
but transformed potential
$K$ for the case(ii) is not always
a solution of the same equation.
Let us spend the rest
of the present subsection
to comment on the
quasi-\Kahler transformation.
Here we consider
the quasi-K\"ahler transformation
in the presence of
$U(1)$ isometry with respect to
$W$.
In this case
the transformation \Inva\ is classified
by the following three cases:
$$
\eqalignno{
&K \rightarrow \Kh= K-(w+\wb)
(\lambda(u)+{\bar \lambda}(\ub)),
&(A.15)\cr
&K \rightarrow
\Kh= K-k(w+\wb),
&(A.16)\cr
&K\rightarrow
\Kh=K-\lambda(u)-
{\bar \lambda}(\ub).
&(A.17)}
$$
Let $K$ be a solution of (A.13).
The potential ${\hat K}$
which is generated from $K$
under the quasi-K\"ahler
transformation
(A.15) satisfys not
eqn.(A.13) but (A.11).
As is mentioned before,
eqn.(A.11) is transformed
to the original equation
under the coordinate
transformation (A.12).
Thus the transformed $K$
is also a solution only
when the quasi-K\"ahler transformation
(A.15) is followed by the coordinate
transformation (A.12).
The K\"ahler transformation
(A.16) make the constant $c_2$
to be shifted by a real constant $-k$.
To obtain the solution of original
equation we perform
the coordinate transformation
$$
e^{k/2}du=dF,\quad e^{k/2}d\ub=d{\bar F}.
\eqno(A.18)
$$
In turn the K\"ahler transformation
(A.17) is invariance of (A.13).

Since a real part of $c_2$ can be
absorbed by the quasi-\Kahler
transformation (A.16)
or the coordinate transformation (A.18),
the imaginary part of $c_2$ is relevant.
If we consider the case $c_2=0$,
the resulting geometry
has $(2,2)$ signature.
In order to have euclidean signature
we must set $c_2=i\pi$.

\centerline{\bf APPENDIX.B}

In this section we consider
duality transformation.
As was explained in ref.[\LR,\GHR]\
this duality
can be described by interchanging
twisted chiral superfields with
chiral ones.
Let us consider the case that the
potential $K$ has one Killing symmetry
with respect to $V$ and
is of the form
$$
K=K(U,\Ub,V+\Vb),
\eqno(B.1)
$$
where $V$ is a twisted chiral field,
whereas $U$ is a chiral field.
We denote the above $U(1)$ isometry
as $U(1)_v$
for simplicity.

The `dual' potential $\Kt$
is obtained as a Legendre
transform of $K$;
$$
\tilde K(U,\Ub,V+\Vb,\Psi+\bar\Psi)=
K(U,\Ub,V+\Vb)-(V+\Vb)(\Psi+\bar\Psi),
\eqno(B.2)
$$
with
$$
{\partial K\over\partial v}-
(\psi+\bar\psi)=0,
\eqno(B.3)
$$
where $\Psi$ is a chiral field.
Since the dual potential $\Kt$
is described by two chiral
superfields, the dual transformation
explained above produces
a torsionless \Kahler manifold.
It follows that the dual metric
has the following form:
$$
\tilde G_{\mu\nu}=
\left(
\matrix{
0&\tilde K_{\psi\psib}&0
&\tilde K_{\psi\vb}\cr
\tilde K_{\psi\psib}&0
&\tilde K_{v\psib}&0\cr
0&\tilde K_{v\psib}&0
&\tilde K_{v\vb}\cr
\tilde K_{\psi\vb}&0
&\tilde K_{v\vb}&0\cr}
\right).
\eqno(B.4)
$$
On the other hand if the Killing symmetry
is with respect to
a chiral superfield $U$,
the corresponding dual metric
has opposite signature to (B.4).

Since we are considering the case
that there are one chiral and one
twisted chiral superfield,
the duality transformation
produces a torsionless K\"ahlar manifold
explained above.
In order that this N=2 preserving duality
transformation by means
of a Legendre transformation coincides
with the usual abelian T-duality
transformation[\BU], the dual dilaton
field must be
$$
2\Phit=2\Phi-\ln2K_{vv}.
\eqno(B.5)
$$
Since the dilaton field is expressed
by eqn.\tisodila\ for the case(ii),
we obtain linear dilaton backgrounds
after the duality transformation
with respect to
$U(1)_w$ isometry.
For the case(i), if there exist a
$U(1)$ Killing symmetry,
the dual dilaton field is constant.
\endpage
\refout
\end